\documentclass[a4paper,twoside]{article}

\usepackage{epsfig}
\usepackage{subfigure}
\usepackage{calc}
\usepackage{amssymb}
\usepackage{amstext} 
\usepackage{amsthm}
\usepackage[table,xcdraw]{xcolor}
\usepackage[fleqn]{amsmath} 
\usepackage{pslatex}
\usepackage{enumitem}
\usepackage{textcomp}
\usepackage{natbib}
\usepackage{array,multirow,graphicx}

\usepackage{SCITEPRESS}     

\begin{document}

\title{Towards Classification of Lightweight Formal Methods}


\pagestyle{myheadings}
\markboth{PREPRINT}{PREPRINT}

\author{ 
\authorname{ 
Anna Zamansky\sup{1},  Maria Spichkova\sup{2}, Guillermo Rodriguez-Navas\sup{3},\\ 
Peter Herrmann\sup{4}, Jan Olaf Blech\sup{5}
}
\affiliation{
\sup{1}IS Department, University of Haifa, Haifa, Israel, annazam@gmail.com
}
\affiliation{ 
\sup{2}School of Science, RMIT University,Melbourne, Australia, maria.spichkova@rmit.edu.au
}
\affiliation{
\sup{3}IDT, M\"{a}lardalen University, V\"{a}ster{\aa}s, Sweden, guillermo.rodriguez-navas@mdh.se
}  
\affiliation{
\sup{4}IIK, Norwegian University of Science and Technology, Trondheim, Norway
}
\affiliation{
\sup{5}BHTC GmbH, Lippstadt, Germany,  joblech@gmail.com
} 
}

\abstract{The use of lightweight formal methods (LFM) for the development of industrial applications has become a major trend.
Although the term ``lightweight formal methods'' has been used for over ten years now, there seems to be no common agreement on what ``lightweight'' actually means, and different communities apply the term in all kinds of ways. 
In this paper, we explore the recent trends in the use of LFM, and establish our opinion that cost-effectiveness is the driving force to deploy LFM.
Further, we propose a simple framework that should help to classify different LFM approaches and to estimate which of them are most cost-effective for a certain software engineering project. We demonstrate our framework using some examples.\\
~ \\ ~
}

\onecolumn 
\maketitle 
\normalsize \vfill

\section{\uppercase{Introduction}}
\label{sec:introduction}
\emph{Lightweight formal methods} (LFM) is an increasingly popular approach to apply \emph{formal methods} (FM) in a rapid and cost-effective way. Already in the 90s, Jackson and Wing admitted that ``by promoting full formalization in expressive languages, formalists have unwittingly guaranteed that the benefits of formalization are thinly spread'', cf. \cite{JWLFM}. To overcome this obstacle, they suggested to use LFM, an approach which emphasizes partiality, and predicted that this is more likely to be applied in practice. 
\cite{jackson2001lightweight} identified four dimensions of partiality, i.e., language, modeling, analysis and composition, that were later characterized in greater detail by \cite{barner2005lightweight}:

\begin{itemize}
\item 
Partiality in language: The idea is that formal specification languages shall be designed in a way that the specifications can be automatically analyzed by tools (as opposed to earlier formal specification approaches such as Z, cf. \cite{Spivey_92}, that often require manual analysis); 
\item 
Partiality in modeling: The focus of an FM application shall be on those aspects of a system for which the analysis merits the cost of formalization (as opposed to formalization for its own sake); 
\item 
Partiality in analysis: The expressiveness of a specification language shall be restricted such that it is decidable and/or tractable which makes the use of automated analysis tools easier; 
\item 
Partiality in composition: This dimension focuses on the goal to use FMs together with (formal or semi-formal) description techniques without providing a precise link between them. Thus, it will be easier to handle the complexity of the specification and analysis process.
\end{itemize}
An approach alternative to the partialities is introduced by \cite{garavel} who associate formal methods with design flows.
The authors make a distinction between conventional design flows in which no FMs are applied, and formal design flows which incorporate the use of FMs.
For the latter, they further distinguish between fully formal flows  in which FMs are used in all development stages, and partially-formal flows, in which one applies FMs only at certain stages of the life cycle.  
The decision in which stages FMs should be used, depends on the following aspects:
\begin{itemize}
\item 
Using FMs should address the most critical development problems such that the profit is highest;
\item 
FMs shall be only used if they can successfully compete with conventional methodologies.
\end{itemize}
As examples for the application of FMs in partially-formal flows, Garavel and Graf name modeling that focuses on requirements, while analysis is often done by rapid verification and validation (V\&V) using tools such as syntax and semantic checkers, static and dynamic analyzers, or model checkers.

Various lightweight or selective approaches to FM are reviewed by  \cite{costeffective} using the term \emph{cost-effective formal methods}. Cost-effectiveness analysis is a form of economic analysis that compares the relative costs and outcomes (effects) of different courses of action with each other, cf. \cite{Robi:93}. 
This method seems better suited to software development than the cost-benefit analysis in which a monetary value is assigned to the measure of effect.
The reason is that typical cost parameters like time usage or human resources can often hardly be translated into a monetary equivalent. 
However, cost and benefit analysis for projects applying FMs is known to be extremely challenging, cf. \cite{stidolph2003managerial,bowen1995seven}.

As there is no proper classification of FMs wrt. lightweightedness,  it is highly challenging to decide based on the available FM literature, which particular LFM should be used in a certain context.

\emph{Contributions:}
We examine the recent trends of research carrying the term ``lightweight formal methods'', and provide a framework for the classification of LFM applications. The idea behind our framework is that we perceive the aspects of partiality \cite{jackson2001lightweight,garavel} only as a symptom of cost-effectiveness, which seems to be the main driving force behind deploying LFM instead of FM.
In particular, we carve out a number of aspects that can be used to characterize the properties of LFM.
Based on that, we present a small framework consisting of four simple questions that one may ask to understand the gains and costs of employing a certain LFM in a software development project.

\section{\uppercase{Related Work}}
\label{sec:related}
 
We restrict ourselves on publications actively using the term ``lightweight formal methods''.
We check out how this term is used and how the approaches are mapped to the dimensions of partiality, according to the approach proposed by  \cite{jackson2001lightweight}.  
We searched for papers using the phrase ``lightweight formal method(s)'' in their title using Google Scholar and found 36 entries. Excluding unpublished papers and works focusing on education, we came up with 14 relevant papers that are published between 2005 and 2017.
They are summarized in Table~\ref{tab:reviewedpapers}.

\begin{table*}[tb]
\centering
\caption{Reviewed papers}
\label{tab:reviewedpapers}
{\small
\begin{tabular}{|l|l|l|m{6.2cm}|l|}\hline
 n&  Reference & FM  & Description & Partiality \\ \hline \hline
 1&  \cite{braga2010transforming} & Alloy & conceptual model validation & modeling/analysis  \\ \hline
 2& \cite{shao2009incremental} & Alloy &   formal verification of partial models of  code & modelling/analysis\\  \hline
 3&  \cite{breen2005experience} & FSM & requirement specification &  language \\ \hline
 4&  \cite{yang2012specification} & Alloy & specification-based test repair & modelling/analysis\\ \hline
 5 & \cite{valles2009use} & Alloy & specification of static aspects & analysis\\ \hline
 6 & \cite{horl2000validating} & VDM & applying test cases to specification & analysis \\ \hline
7 & \cite{gilliam2005application} & SPIN& model checking of system components & modeling\\ \hline
8 & \cite{bontemps2005lightweight} & LSC & Scenario-based specification & analysis  \\ \hline
9 & \cite{boyatt2007lightweight} & Alloy & partial specification and analysis & modelling/analysis\\ \hline
10 & \cite{valles2012using} & Alloy & consistency checking for class/object diagrams & analysis\\ \hline
11 & \cite{oda2016viennatalk} & VDM & flexible modeling/coding with rigorous checking& analysis \\ \hline
12 & \cite{slaymaker2010formalising} & Alloy & consistency checking of Z models& modelling/analysis \\ \hline
13 & \cite{hao2016designing} & Alloy & Designing normative systems& modeling/analysis \\ \hline
14  & \cite{radar_icse2017} & RADAR  & Requirements and architecture decision analysis & modeling/analysis\\
\hline
\end{tabular}
}
\end{table*}

The majority of these papers apply formal methods based on the specification language Alloy, cf. \cite{jackson2012software}. 
Alloy is often called ``lightweight'' which is no wonder since it is developed by one of the original inventors of the term ``lightweight formal methods''.
\cite{braga2010transforming} proposed
an approach to facilitate the validation process of conceptual
models defined in the semi-formal conceptual modeling language OntoUML by transforming these models
into specifications in the formal language Alloy and
using its analyzer to generate instances of the model and assertion counter-examples. 
A similar rationale is provided in \cite{yang2012specification}, where Alloy is employed for specification-based repair. 
\cite{shao2009incremental} proposed an optimization of scope-bounded checking of Alloy specifications. 
All of the above works share one thing in common: The only context in which the word ``lightweight'' is mentioned is the reference to Alloy.\footnote{Only \cite{braga2010transforming} refer also to the original rationale of \cite{jackson2012software}.}

In his book introducing Alloy, however, Jackson writes: ``I sometimes call my approach `lightweight formal methods' because it tries to obtain the benefits of traditional formal methods at lower cost, and without requiring a big initial investment. Models are developed incrementally, driven by the modeler’s perception of which aspects of the software matter most, and of where the greatest risks lie, and automated tools are exploited to find flaws as early as possible'', cf. \cite{jackson2012software}.
According to this definition, the use of Alloy may be mapped to the partialities in \emph{modeling} and \emph{analysis} according to \cite{jackson2001lightweight}. 

A definition of the lightweightedness of Alloy is provided by Valles-Barajas who sees a central purpose in the exposition of flaws instead of full verification: 
``The lightweight methodology may seem to go against the grain of the traditional
approach to formal methods. However, a more limited scope can be very effective
at, for example, providing counterexamples which expose design flaws. It does not
promise full verification or full system coverage. One notable lightweight approach
is Alloy", cf. \cite{valles2009use}. This refers to  partiality in {\em analysis}.  

\cite{breen2005experience} applied finite state machines to model commercial embedded system product lines providing a different meaning to the term ``lightweight formal methods'':
``Specifiers
should be able to enjoy formal method benefits
such as lack of ambiguity and amenability to automated
checks but without needing to learn arcane mathematical
notations or subtle theoretical concepts''. 
In contrast to the previous definitions, this can be mapped to partiality in \emph{language}. 

\cite{bontemps2005lightweight} proposed a method to model check scenario-based models using live sequence charts (LSC), introduced by \cite{DaHa:01}. 
Under lightweightedness, the authors understand to give up the completeness property of the model checking procedure. This can be mapped to partiality in \emph{analysis}. 

\cite{gilliam2005application} proposed a framework for system verification using Promela and the model checker SPIN, introduced by \cite{Holz:SPIN04}.
They use a compositional approach that allows one to conduct proofs for a subset of the overall environment in a manner that those results can be extrapolated to the environment at large.
In particular, their approach narrows the focus to components for which security properties have been
identified and which can be modeled.
This approach can be mapped to partiality in \emph{modeling}. 

The remaining two works described applications of variations of the Vienna Development Method (VDM), cf. \cite{BjJo:VDM78}. 
Horl and Aichernig describe an industrial project, in which VDM++ has
been applied to specify a safety critical voice communication system for
air-traffic control. By ``lightweightedness'' the authors define ``taking the advantages of a precise and unambiguous
specification language to raise the quality of a system’s specification, without
focusing on proofs'', cf. \cite{horl2000validating}.
Thus, they use a formal method just to raise quality of the system description but not to prove correctness. This does not fit to well to the partialities of \cite{jackson2001lightweight}.
The best fit seems to be partiality in \emph{analysis} since the avoidance of ambiguity errors seems to be a matter of ``tractability''.
 
\cite{oda2016viennatalk} demonstrated an integrated development environment supporting formal methods in the spirit of supporting
dynamic programming languages, using VDM. By ``lightweightedness'', the authors refer to the combination of flexible live modeling resp. coding with the support of rigorous checking. This can be also mapped to partiality in \emph{analysis}. 

\cite{radar_icse2017} presented RADAR, a 
modeling language that is a simplified form of quantitative goal
models designed to be similar to simple equations that software
architects use for back-of-the-envelope calculations.

To summarize, Alloy is the most popular LFM in the reviewed publications (7 out of 14 papers).
Moreover, it seems common to use the term ``lightweight formal method'' in the context of Alloy without providing further rationale why this term is used at all.
The most popular dimensions of partiality according to \cite{jackson2001lightweight} are \emph{modeling} and \emph{analysis}.

\begin{figure*}[ht!]
\caption{V-model}
\begin{center}
\centering
\includegraphics[scale=0.39]{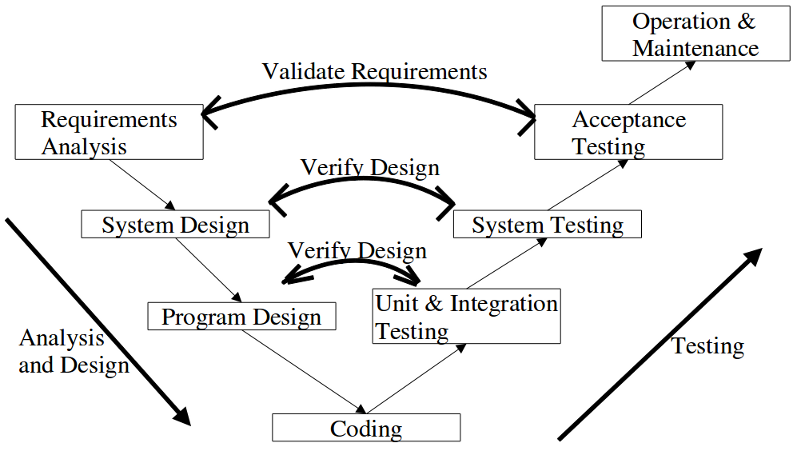}
\end{center}
\label{fig:v-model}
\end{figure*}

The dimensions of partiality \cite{jackson2001lightweight,barner2005lightweight} are only symptoms in our view, and it seems essential to consider cost-effectiveness, in which we see the real driver behind deploying LFM. 
As highlighted by Easterbrook et al., ``lightweight application of formal methods can offer a cost-effective way of
improving the quality of software specifications'', cf. \cite{easterbrook1998experiences}. 
Therefore, to understand how LFMs affect cost-effectiveness, formal methods must be put in the context of the development process.

A plethora of methodologies for software development have been proposed,
and there is a huge amount of literature discussing 
them. All these methodologies aim at developing products reliably and timely,
which differ in several aspects, e.g., design organization, implementation, quality and revision steps, cf.~\cite{ruparelia2010software}.
On one side of the spectrum lies the methodology known as the V-model, proposed by~\cite{FoMo:91}. This methodology can be regarded as an extension of the waterfall model introduced by \cite{Royc:70} and provides a divide and conquer-based development approaches, see Figure~\ref{fig:v-model}. 
The left leg of the ``V'' represents the design activities, whereas the right one stands for the testing activities which also can be used for more in-depth analysis like verification or validation. While we descent the left leg of the ``V'', components and their requirements are identified. Components are then refined, implemented and tested in parallel, before they get integrated again when climbing up the right leg of the ``V''. The divide and conquer-based approach is beneficial to outsourcing the activities in order to separate subcontractors. The requirements for the individual components represent a kind of contract, which can serve as a basis for the application of LFMs.
This organization of the work maps very well the rationale of Model-Based Development (MBD).

The V-model has been criticized for being ambiguous and, like other approaches based on the waterfall model, not capable of handling changes well. It assumes that all relevant requirements about the future product can be collected before starting the design and the implementation. This can be unfeasible for some projects in practice. Testing is supposed to happen at the component level and on the system level. A known risk of the V-model is, however, that system testing activities are often postponed until the last phase of the project. Yet, that increases the probability of failing the whole project since design errors in the early phases might have lead to grave faults.
Solving these problems is usually expensive and leads to significant project delays.

Nevertheless, the V-model can be seen as a good reference model from a managerial perspective and is still recommended in many industrial domains. For instance, it is the suggested methodology in the ISO 26262 standard for functional safety of electrical/electronic systems in road vehicles, cf. \cite{ISO26262}. The W (or double vee) model, suggested by \cite{Spil:00}, is an evolution of the V model which clearly specifies the testing activities that should be performed in every step, so errors can be detected and fixed in the step in which they are introduced.

In another approach, the Activity Model introduced by \cite{BrHa:93}, the different phases of the software development process (i.e., requirement engineering, functional design, implementation design (considering non-functional properties), and implementation) varying models are created.
This allows one to analyze the different models separately such that design errors are detected and corrected early.
Further, the models can be used as milestones for the interaction between the software developer and the remitter who has to accept the models for a phase, before the next one can be started. 

On the other side of the spectrum, there are methods for Rapid Application Development, with the Agile methodology being the most extreme exponent, cf.~\cite{Coll:11}. 
Agile development emphasizes teams and human interaction over processes and documentation. Some characteristics of the Agile life cycle are a very lean start-off design and analysis phase, iterative design and development, incremental design and development, parallelization of design and development with frequent synchronization and integration, and documentation reduced to a minimum. In fact, the team only documents what is absolutely necessary. This should not be taken as a total lack of documentation. Some artifacts, like use cases, lists or diagrams can be exchanged among developers and testers frequently.

In most of the approaches named, software design methodologies do neither specify whether FM have to be applied at all, nor at what points in the engineering process they should be used. Due to its generality, however, developers of FM often describe their tools in relation to the V-model. One FM is typically applicable within a particular phase (e.g. sanity checking to analyze the requirement specification, model checking and theorem proving to verify the design, code analyzers to verify the implementation). Some formal methods connect different phases, such as model-based testing, which uses the specification in order to derive the optimal test set with respect to a certain property.

A vast amount of FMs exist that can be applied to the various phases of the software engineering process like the specification, verification and validation of requirements, functional system design, non-functional system design, software design, implementation of code. Other FMs support the various testing disciplines like unit or integration testing.

It seems that FMs are rarely used in Agile testing.
Probably, the complexity of deploying FM is seen as unsuitable to its idea of lean development and analysis.
One might wonder whether the application of LFMs might bring here a change since the concept of lightweightness seems to be a good fit to the idea of agility.
Yet, up to now very little has been published about the application of LFMs in Agile development.

\section{\uppercase{Classification framework}}
\label{sec:framework}

A typical application of an FM is the construction of a formal model specifying important aspects of the system to be developed (e.g., requirements) combined with a formal analysis of this model (e.g., by verification, validation, type or model checking).

Both, the creation of specifications and their analysis require resources, and their extent may highly depend on the stage in the development cycle, the used techniques, and the characteristics of the project.
It can be very complex to develop a real-life system such that specifying and analyzing it may consume a lot of time which is definitely a cost-factor.
Moreover, specifying and analyzing subtle functionality resp. quality properties (e.g., robustness or realtime) can be quite complicated.
In this case, one often has to fall back on the skills of experts in the field which can be a significant cost-factor as well.
Both cost-factors depend heavily on the kind of modelling and analysis approaches to be used, and not all techniques fit to all systems to be developed.

On the other side, we have also to consider the gains achieved by the usually higher quality, a system has when it has been engineered using FMs.
Since many design errors are detected in the specification and analysis process, the software should be deployable with significantly lower maintenance and compensation costs.
Of course, also the gains depend on the FMs used, in particular, on the rigor of the analyses conducted by them. 

Offsetting the costs and gains of using FMs leads to the following trade-off with respect to deploying LFMs:
Using LFMs with increased partialities according to \cite{jackson2001lightweight} will likely reduce the cost-factors of time and expert use.
On the other side, the gains due to a better quality of the produced software tends also to be lower.
The trade-off is of a qualitative nature which brings us to the problem that qualitatively analyzing costs associated with FM applications in projects is very challenging and understudied, cf. \cite{stidolph2003managerial}.

Despite these challenges, the managers of a software development venture have to decide which FMs to apply and in which phase of the engineering process.
We do not claim to provide a complete framework that may guide this decision.
In general, software is diverse, such that there is highly varying requirements to be fulfilled when creating it.
This makes it hard to map the ever growing number of available FMs successfully to these requirements in order to offer the most cost-effective solution.

In the following, we mention a number of central issues that we consider suitable to guide the selection of FMs based on cost-effectiveness.
Based on these aspects, we thereafter define four simple questions that can provide some support for the classification of LFMs and, in consequence, for the decision, which one to use in a certain situation. 

\begin{itemize}
\item 
FMs can be applied at different development stages, from requirement engineering via the functional and deployment design phases towards the implementation, cf.  \cite{BrHa:93}.
Several works highlight the benefits of early application of FMs to formalize and validate requirements, e.g., \cite{berry2002formal}. 
In this way, errors that are introduced in the early design steps, can be detected and corrected before they become costly.
Furthermore, in early design phases system descriptions are relatively abstract which makes both the specification and formal analysis of the systems easier.
Other work supports the application of
formal methods at the end of the life cycle, i.e., directly on the source code
of the software, cf. \cite{shao2010certified}. One argument supporting this view is that even if many programming languages lack a formal semantics and any of their low-level
features are implementation-dependent and underspecified,
the late design artifacts expressed in these languages are often
the most precise and unambiguous descriptions of a system, especially
when compared to informal artifacts produced during earlier design
steps.
\item 
An application of a FM can focus on different parts and aspects of the system. For example, the most complex parts of the system, or the most critical safety or security properties can be formalized.
Another reason to treat certain parts of a system formally, is that some properties like performance issues can hardly be described in an informal or semi-formal way.
Beyond that, sometimes the developers have to obey regulations for licensing the software that determines this question, cf. e.g., \cite{Lawl:74}.
\item 
There is a wide variety of formal languages that can be used in a FM application, and we already stated that the selection can have a significant impact on the resources to be used. In particular, the tool support is often critical.
On the one hand, some tools (e.g., model checkers) work highly automatically while others (e.g., theorem provers) are more interactive and need a higher degree of human intervention.
On the other one, many tools demand a significant learning curve such that familiarity and experience of their users get a major factor.
\item 
Due to the varying complexity of the software to be developed and the FMs to be used, we are often restricted in the decision, whom to schedule certain tasks. 
Some FM applications require highly trained mathematicians or practitioners with hard-to-acquire expertise, cf. \cite{hall1990seven}.
In contrast, other techniques are much easier to handle or even hide formal modeling and analysis to their users, cf. e.g., \cite{KrSH:JSS09}.
\end{itemize}

To describe the aspects mentioned above in a more striking form, we summarize them in the form of four relatively simple questions that, however, can be helpful to decide about the cost-effective use of FM:\\
\emph{WHEN:} At which development stage should FMs be applied?\\ 
\emph{WHAT:} For what parts/aspects of the system should FMs be applied?\\ 
\emph{HOW:} How rigorous shall the modeling and analysis be and what languages and tools should one use to achieve that?\\
\emph{WHO:} Which human resources should be deployed?

In our opinion, envisioning these questions for possible LFM candidates facilitates to foresee the advantages and disadvantages of them for a certain development project and in a particular situation (e.g., the availability of experts).
This should make it easier to give relatively fair predictions about the factual cost-effectiveness of the different methods such that their benefits can be compared, and meaningful managerial decisions can be taken.

\section{\uppercase{Classification Examples}}
\label{sec:examples}

To investigate the above claim, we apply the proposed framework on a number of  LFMs actively applied in the FM community,
and characterize their cost-effectiveness by answering the four questions for each of them.

\subsection{Software Specification in cTLA}
The \emph{compositional Temporal Logic of Actions} (cTLA)  
is a variant of Lamport's Temporal Logic of Actions (TLA) introduced by \cite{Lamp:02}.
It enables us to describe components and constraints of systems as separate specifications in a process-oriented way that can be composed by defining the joint execution of actions.

Process composition in cTLA has the character of superposition, cf. \cite{BaKS:89}, i.e., a property of a single process is also a property of each system containing it.
This makes it possible to reduce the proof that a system fulfills a property to the verification that an (often very small) subsystem fulfills this property. 
These ingredients allow for the development of specification frameworks that contain libraries of cTLA process types that model typical functions of a certain application domain.
Moreover, a framework contains libraries of theorems stating that, under some side conditions, a cTLA process or a subsystem of such processes fulfill a more coarse-grained functionality or a property.
The theorems were verified by the framework developers such that the user needs only to check that their side conditions hold and that the processes in the fine-grained systems are consistently coupled.
These checks are usually simple and can also be automated, cf. \cite{HKDG:02}. 
Such frameworks were developed for the service verification of communication protocols,  
hazard analysis of chemical plants,  
and security policy verification of component-structured software, cf. e.g. \cite{Herr:FORTE03}.

For this approach, we can answer the four questions in the following way:
\\
\emph{WHEN:} The approach can be used at the requirement engineering (RE) and the functional design stage of the software development cycle.\\
\emph{WHAT:} A framework can be used to specify and analyze functions that are supported by its libraries.\\
\emph{HOW:} The approach is formally rigorous since it is based on TLA and one should use the checking tool delivered by the approach.\\
\emph{WHO:} Humans applying this approach should know the basics of TLA and cTLA specification and verification. 
Thanks to the clear structure and the tools available, however, they do need in-depth knowledge in verification based on temporal logic.

In consequence, we can expect the approach to be very cost-effective as long as there is a framework available for the particular problem since one gets a high degree of formal rigor without having to carry out complex verifications.
For problems, not yet covered by a framework, however, the approach will have no advantage over traditional TLA.

\subsection{Software Development with Reactive Blocks}

\emph{Reactive Blocks} is a tool-set
that also facilitates the reuse of functionality but on a much lower level of the software engineering cycle, cf. \cite{KrSH:JSS09}.
Basically, it is a MBD technique for reactive systems based on the programming language Java.
The reuse of functionality is supported by special \emph{building blocks}, which may contain recurring functionality that can be  reused in several software projects.
To facilitate the reuse further, the interface behavior of each building block is specified by an \emph{External State Machine (ESM)}.
The building blocks can be stored for further use in special libraries of the tool. 
The behavior of a building block is specified using UML activities: %
a graph structure that resembles Petri nets and facilitates the understanding of control and data flows in a software.
The nodes represent relevant software elements like building blocks, Java methods, or timers as well as elements to coordinate flows, e.g., forks, merges.

Since the OMG did not provide the activities with a formal semantics, one was defined by the developers
\cite{KrHe:FORTE10}.
Based on that, a model checker could be integrated into the Reactive Blocks tool-set \cite{KrSH:JSS09}, which allows the verification of various correctness properties, e.g., that both, a building block and the environment using it preserve its ESM.
The model checker is completely transparent to the user of the tool who simply has to push an {\tt Analyze} button.
Thus, the approach follows the ``Disappearing Formal Methods'' concept proposed by \cite{Rush:00} according to which the formal issues of the verification process shall be hidden to the user of the tool. 

The four questions can be answered as follows:
\\
\emph{WHEN:} The Reactive Blocks tool-set covers the whole software engineering process except for requirements engineering.\\
\emph{WHAT:} It covers the whole system design.\\
\emph{HOW:} While the whole system is modeled, the formalism is not very rigorous.
The syntax facilitates the understanding of the various control and data flows of a system which is helpful in complex systems with several coordinated threads.
The built-in model checker, however, can check only relative abstract properties of the flows, e.g., compliance with the ESMs or freedom of deadlocks. \\
\emph{WHO:} After a short training period, the tool can be used by everybody who is able to program in Java. 

Altogether, we have a formal method that is comprehensive and very helpful to engineer systems with complicated control and data flows.
It is cost-effective thanks to the reuse factor and the easiness to apply the model checker.
On the other side, the gain achieved by formal analysis is limited. 
For instance, errors in the code of the operations cannot be detected by the model checker such that normal code testing is still necessary.

\subsection{Software Development with AF3}

\emph{AutoFocus3} (AF3) is a tool for the model-based development of distributed, reactive, and embedded software systems,  
cf. \cite{aravantinos2015autofocus,holzl201013}.  
The tool supports formal verification, automatic test-case generation, simulation of the system by test-cases, and on the fly consistency checks. 
The earlier versions of the tool were successfully applied in a number of scientific and industrial case studies \cite{DentumKeylessEntry,feilkas2009top,broy2008service,VerisoftXT_FMDS,verisoftxt_praxis,holzl2010safety,spichkova2013we,dobi2013model},
providing a basis for the further development of the tool as well as for expanding the coverage of the software development phases. 
On the verification level, AF3 provides several options that can also be combined, for example:
\begin{itemize}
\item Model-checking support, covering a large subset of the AF3 language.
\item Easy to write verification conditions (via patterns or sequence diagrams).
\item Check if the guarantees of neighboring components fulfill the assumptions of a component.
\item Analysis whether an application realizes a scenario specified by a Message Sequence Chart diagram.
\item Lightweighted analyses for possible non-determinism in state automata.
\end{itemize}
Recent research work 
focuses also on an automated 
generation of the documentation from the model, to increase the understandability of the AF3 models, cf. \cite{vo2016model}. 

For this approach, we can answer the four questions as follows:
\\
\emph{WHEN:} AF3 can be used at all stages of the development (RE, design of the logical architecture, the deployment and the scheduling, hardware architecture, etc).\\
\emph{WHAT:} AF3 covers the whole system design.\\
\emph{HOW:} The approach is not really rigorous. The overall goal is to hide the formal aspects of modeling and verification from the developer behind the tool support, including visual modeling, tables and natural language specification.\\
\emph{WHO:} The tool is easy to use, as it allows specification using state automata, source code, or tables. No specific programming skills are required.

The advantage of this approach is that it covers all phases of the software development as well as the high usability of the AF3 tool.

\section{\uppercase{Conclusions}} 
\label{sec:conclusions}
In this paper, we make a step towards creating a ``map of the jungle'' of lightweight formal methods, which will enable researchers and practitioners to compare between different applications of LFMs and choose the LFM most appropriate for their particular case more easily. To this end, we were carrying out a literature review of recent works on LFM applications, exploring the way in which the term is used and what it is supposed to mean. Our review reveals that the most popular applications of LFMs is using Alloy, and that usually no rationale behind its lightweightedness is provided. The dimension of partiality in composition is not addressed in the papers reviewed.
This contradicts with our view that the integration of LFMs into the development process should be a key issue. We further formulated a small framework including four basic questions in order to facilitate the classification of various LFM applications. By considering LFMs in the setting of the development process, the answers to these questions should provide better orientation towards cost-effectiveness.

\bibliographystyle{apalike}
{\small

}

\vfill
\end{document}